\documentstyle[graphicx,prl,aps]{revtex}

\begin{document}

%\draft

\title{Magnetic Behavior of {La$_7$Ru$_3$O$_{18}$}}

 \setlength{\baselineskip}{20pt}
 \setlength{\topmargin}{-0.25in}
%\renewcommand{\baselineskip{2}}

%%%%%%%%%%%%%%%%%%%%%%%%%%%%%%%  authors

\author{P. Khalifah}
\address{Department of Chemistry and Princeton Materials Institute, Princeton University, Princeton, New Jersey 08540}
\author{D. A. Huse}
\address{Department of Physics, Princeton University, Princeton, New Jersey 08540}
\author{R. J. Cava}
\address{Department of Chemistry and Princeton Materials Institute, Princeton University, Princeton, New Jersey 08540}

\date{\today}

\maketitle

%%%%%%%%%%%%%%%%%%%%%%%%%%%%%%%  abstract

%\pacs{75.30 Kz, 75.50 Ee}

% 75.30 Kz = Magnetic phase boundaries (including magnetic transitions, metamagnetism, etc.)
% 75.50 Ee = Antiferromagnetics

\begin{abstract}

Rhombohedral La$_7$Ru$_3$O$_{18}$ can be considered to be nearly
geometrically frustrated due to its close structural similarity to
the strongly geometrically frustrated compound
La$_{4.87}$Ru$_2$O$_{12}$.   The magnetic ordering of
La$_7$Ru$_3$O$_{18}$ was explored using a combination of dc and ac
magnetic susceptibility measurements. The magnetic phase diagram
shows two different ordering regimes. A low field magnetic phase
occurs at applied fields of 0-3T and temperatures below 10K, while
a high field phase is found at fields of 3-5T and temperatures
below 9K.

\end{abstract}

%Note that little mention of the relation to geometric frustration is made...

%%%%%%%%%%%%%%%%%%%%%%%%%%%%%%%  intro

%PACS: 75.30 Kz, 75.50 Ee

%\twocolumn \narrowtext

\section{Introduction}

The long range ordering of a compound with antiferromagnetic
interactions between spins can be hindered by certain geometric
arrangements of the magnetic ions \cite{ramirez94}.  This
phenomena of geometric frustration is known to occur for a number
of different structures with magnetic ions patterned in specific
highly symmetric arrangements in two or three dimensions.  For
two-dimensional structure types, both triangular and kagom\'{e}
lattices of magnetic atoms can lead to geometric frustration. Good
examples of frustration due to triangular lattices can be found in
delafossite-type compounds such as LiCrO$_2$ \cite{tauber72} and
NaTiO$_2$ \cite{hirakawa85}, while frustration has been
extensively studied in kagom\'{e} lattice compounds such as
SrCr$_8$Ga$_4$O$_{12}$ \cite{schiffer96} which crystallize in the
magnetoplumbite structure.  Although a greater degree of lattice
complexity has been observed for three-dimensional geometrically
frustrated antiferromagnets, the basic building blocks of these 3D
lattices are triangles and tetrahedra.  Some compounds that are
frustrated because of their three-dimensional lattice are the
garnet Gd$_3$Ga$_5$O$_{12}$ \cite{hov80}, the spinel ZnCr$_2$O$_4$
\cite{baltzer66}, the FCC compound K$_2$IrCl$_6$ \cite{cooke59},
and the pyrochlore Dy$_2$Ti$_2$O$_7$ \cite{ramirez00}.

\begin{figure} \begin{center}
\includegraphics[width=7in]{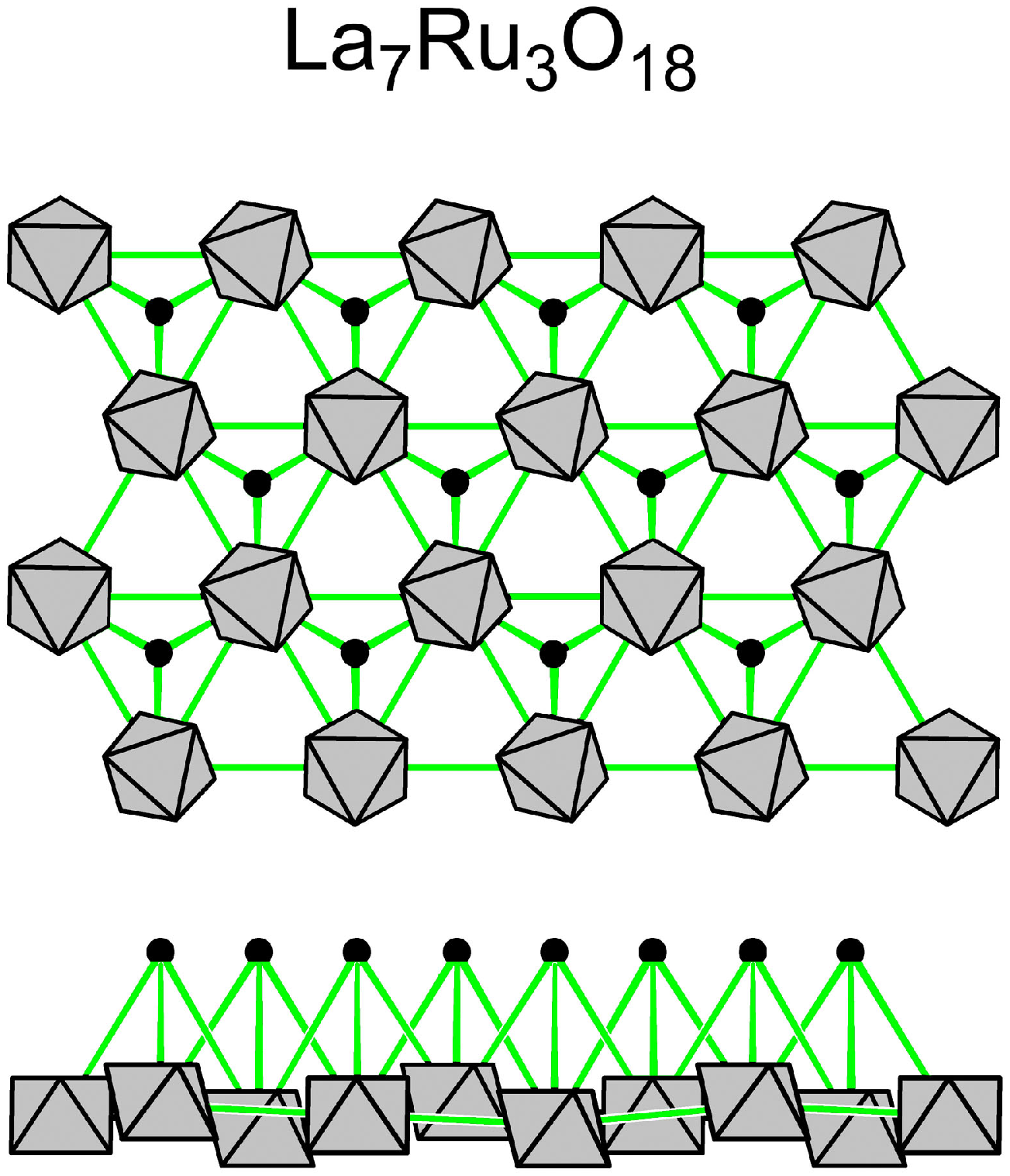}
\caption{Top: View perpendicular to a plane of RuO$_6$ octahedra
(gray) showing the triangular arrangement of Ru atoms in two
dimensions. Ru atoms are at the center of the RuO$_6$ octahedra
while O atoms are at the vertices. Ru atoms in the next higher
plane (shown in black) cap the RuO$_6$ octahedra, giving rise to a
tetrahedral arrangement of Ru sites in three dimensions.  Bottom:
in-plane view, highlighting the tetrahedral arrangement.}
 \label{RUTETRA}
\end{center} \end{figure}

La$_7$Ru$_3$O$_{18}$ has a rhombohedral structure closely related
to that of monoclinic La$_{4.87}$Ru$_2$O$_{12}$, a compound
previously reported to be a strongly geometrically frustrated
antiferromagnet \cite{mypaper3}. Despite the structural
similarities, La$_7$Ru$_3$O$_{18}$ is at most only moderately
frustrated, with a frustration index of $f$ = -$\theta/T_N$
$\cong$ 6. Ru atoms in the $ab$ plane of both compounds are
patterned in a triangular lattice. Each in-plane triangle of Ru
atoms is capped by a fourth Ru atom in a neighboring plane, giving
rise to a tetrahedral arrangement of Ru atoms in three dimensions,
as illustrated in fig. \ref{RUTETRA}. There are three distinct Ru
sites in the crystal structure.  La$_7$Ru$_3$O$_{18}$ was
previously found to have a moment of 3.49 $\mu_B$, and a
Curie-Weiss constant of -58K which indicates medium strength
antiferromagnetic interactions between spins\cite{mypaper3}. In
the course of that initial investigation of La$_7$Ru$_3$O$_{18}$,
behavior more complex than that expected for a simple
antiferromagnet was observed. Here we report a more detailed study
of the magnetic behavior of La$_7$Ru$_3$O$_{18}$, and show that it
has an unusual magnetic phase diagram at low temperatures.

% Talk about physics of geometrically frustrated magnets. and interest
% in having systems that are nearly frustrated...

%Geometrically frustrated antiferromagnets (GFAs) will have an
%ordering transition when cooled to sufficiently low temperatures.
%The type of ordering has been correlated to the nature of the
%deviation from Curie-Weiss behavior that occurs upon cooling
%(cite). GFAs that have enhanced magnetism (relative to the
%Curie-Weiss prediction) have long range antiferromagnetic order
%below the transition temperature, while those that have reduced
%magnetism have spin glass behavior below the transition
%temperature.

%%%%%%%%%%%%%%%%%%%%%%%%%%%%%%%  experimental

\section{Experimental}

% Samples 105-11E and 123-2D were used for magnetic measurments
% T_true = (T_set - 25) in the small box furnaces

Starting materials were dried La$_2$O$_3$ (99.99\%, Alfa) and
RuO$_2$ (99.9\%, Cerac). La$_7$Ru$_3$O$_{18}$ was made by mixing
La$_2$O$_3$ and RuO$_2$ in the stoichiometric ratio or with a
slight Ru excess. The powders were placed in dense alumina
crucibles and heated at 775, 850, and 875 $^\circ$C for at least
two days each with multiple intermediate grindings. Samples were
then annealed for about two weeks at 875 $^\circ$C with multiple
regrindings until judged to be single phase by powder x-ray
diffraction.

Both ac and dc magnetic susceptibility measurements were performed
on the Physical Property Measurement System (PPMS, Quantum
design). Field cooled (FC) measurements were done by applying the
field, cooling the sample, and then measuring the temperature
dependence of the magnetic susceptibility upon heating. Field
sweeps were done by ramping the field from 9T to -9T to 9T. The ac
susceptibility measurements were done at a frequency of 10kHz with
an amplitude of 12Oe. All ac susceptibility measurements made in
the presence of a constant dc field were done with field cooling.

%%%%%%%%%%%%%%%%%%%%%%%%%%%%%%%  results

\section{Results and Discussion}

\begin{figure} \begin{center}
\includegraphics[width=7in]{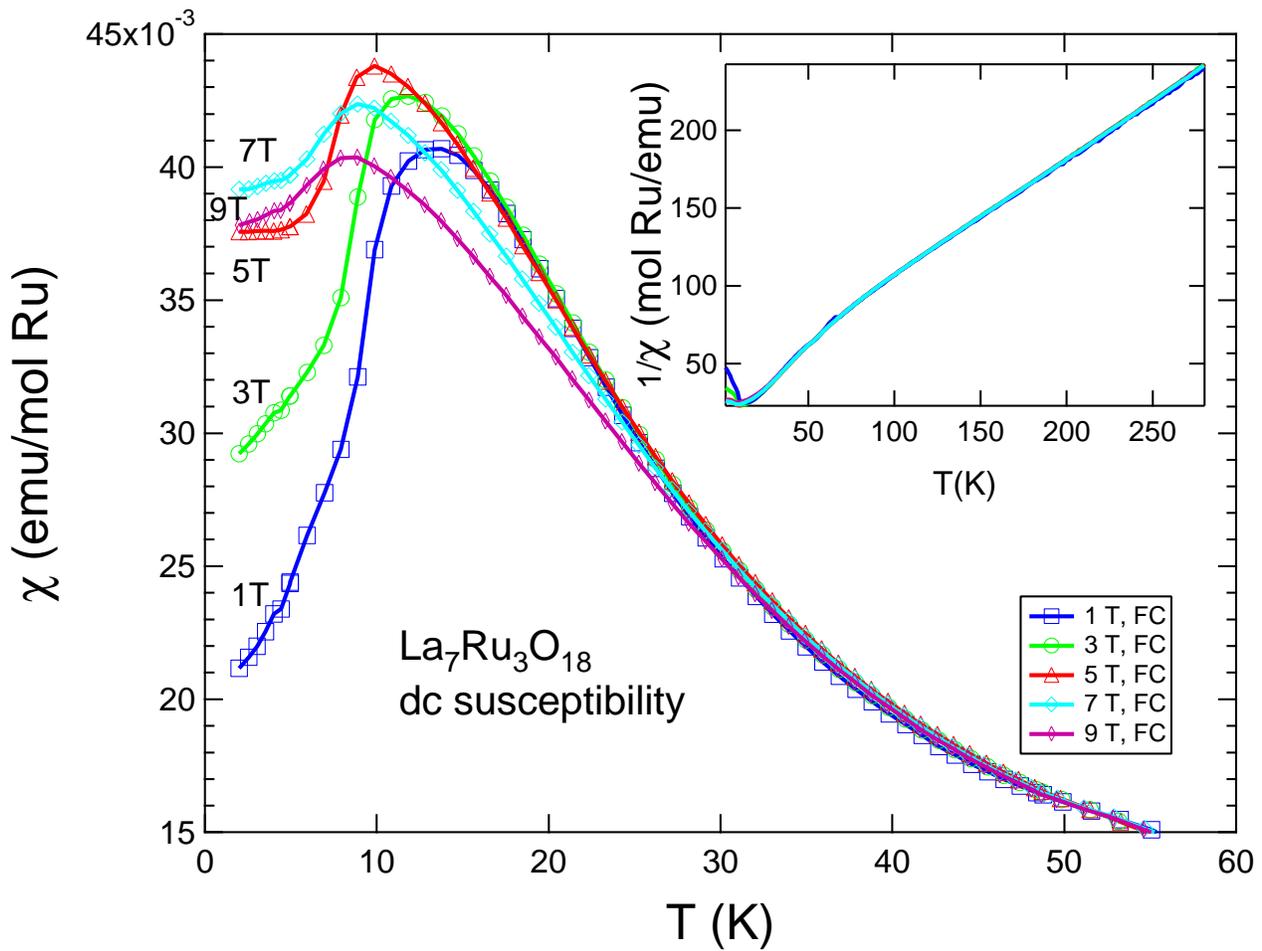}
\caption{Magnetic susceptibility ($\chi_{dc}$ = M/H) at low
temperatures. Inset shows linearity and field-independence of
1/$\chi_{dc}$ at higher temperatures.  Data for H=1T, 3T, 5T, 7T,
and 9T are superimposed.}
 \label{CHIDC}
\end{center} \end{figure}

La$_7$Ru$_3$O$_{18}$ obeys the Curie-Weiss behavior at
temperatures above 60K for all applied fields studied (0-9T), as
seen in the inset to fig. \ref{CHIDC}. There is no difference
between the field cooled and zero-field cooled data (not shown),
indicating that the low temperature drop in the magnetic
susceptibility (fig. \ref{CHIDC}, main panel) represents a
transition to a long range ordered antiferromagnetic ground state
rather than a spin glass ground state.  Although the manner in
which the field was applied does not affect the data, the
magnitude of the field does. The maximum in the absolute
susceptibility ($\chi_{dc}$ = M/H) is at 14K at an applied field
of 1T, and drops to 9K as the applied field is increased to 9T.
Furthermore, the magnitude of the low temperature dc
susceptibility ($\chi_{dc}$ = M/H) increases with increasing
applied field in the range of 0 $<$ H $<$ 7T.  For
antiferromagnets, increasing the applied field has no effect on
$\chi_{dc}$ until sufficiently high fields are reached, leading to
saturation of the spins and causing $\chi_{dc}$ to decrease with
further increases in H. Since there is no simple explanation for
the field behavior of $\chi_{dc}$, it seems that
La$_7$Ru$_3$O$_{18}$ may have multiple types of magnetic ordering.

\begin{figure} \begin{center}
\includegraphics[width=7in]{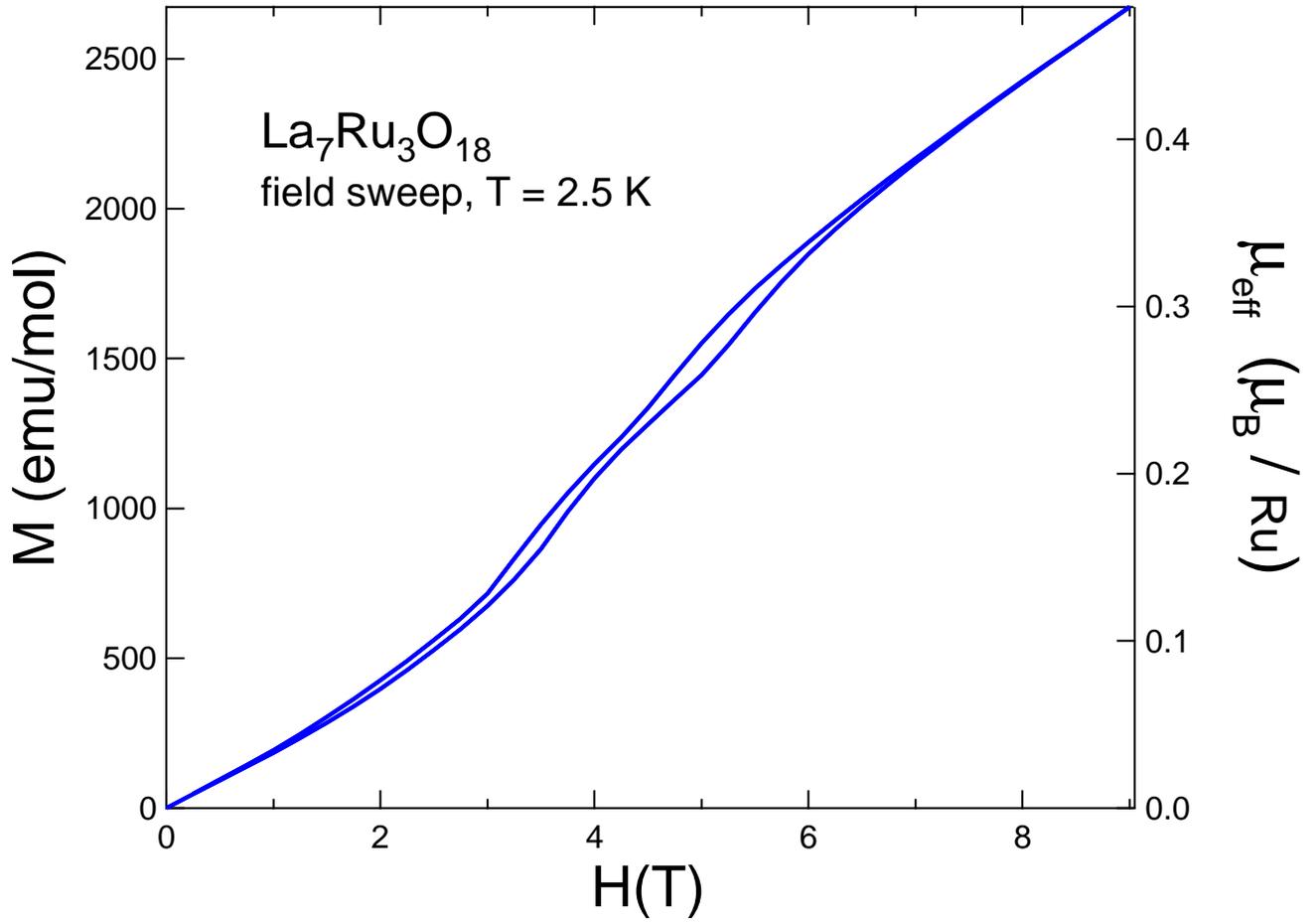}
\caption{Field dependence of magnetization at $T$ = 2.5K showing
the two distinct regions of hysteresis. Only the data for H $>$ 0
are shown.}
 \label{MVSH}
\end{center} \end{figure}

In order to better understand the atypical field behavior of the
low temperature susceptibility, field sweep measurements were made
at a variety of temperatures.  Below 15K, two distinct features
were present in these M vs. H loops, and at 5K and below,
hysteresis was observed.  M vs. H data collected at 2.5K is shown
in fig. \ref{MVSH}, with hysteresis clearly visible in both the
low field ($\sim$3T) and high field ($\sim$5T) regimes.  It is
therefore apparent that two distinct magnetic phase transitions
can be induced in La$_7$Ru$_3$O$_{18}$ by applying a magnetic
field.

\begin{figure} \begin{center}
\includegraphics[width=7in]{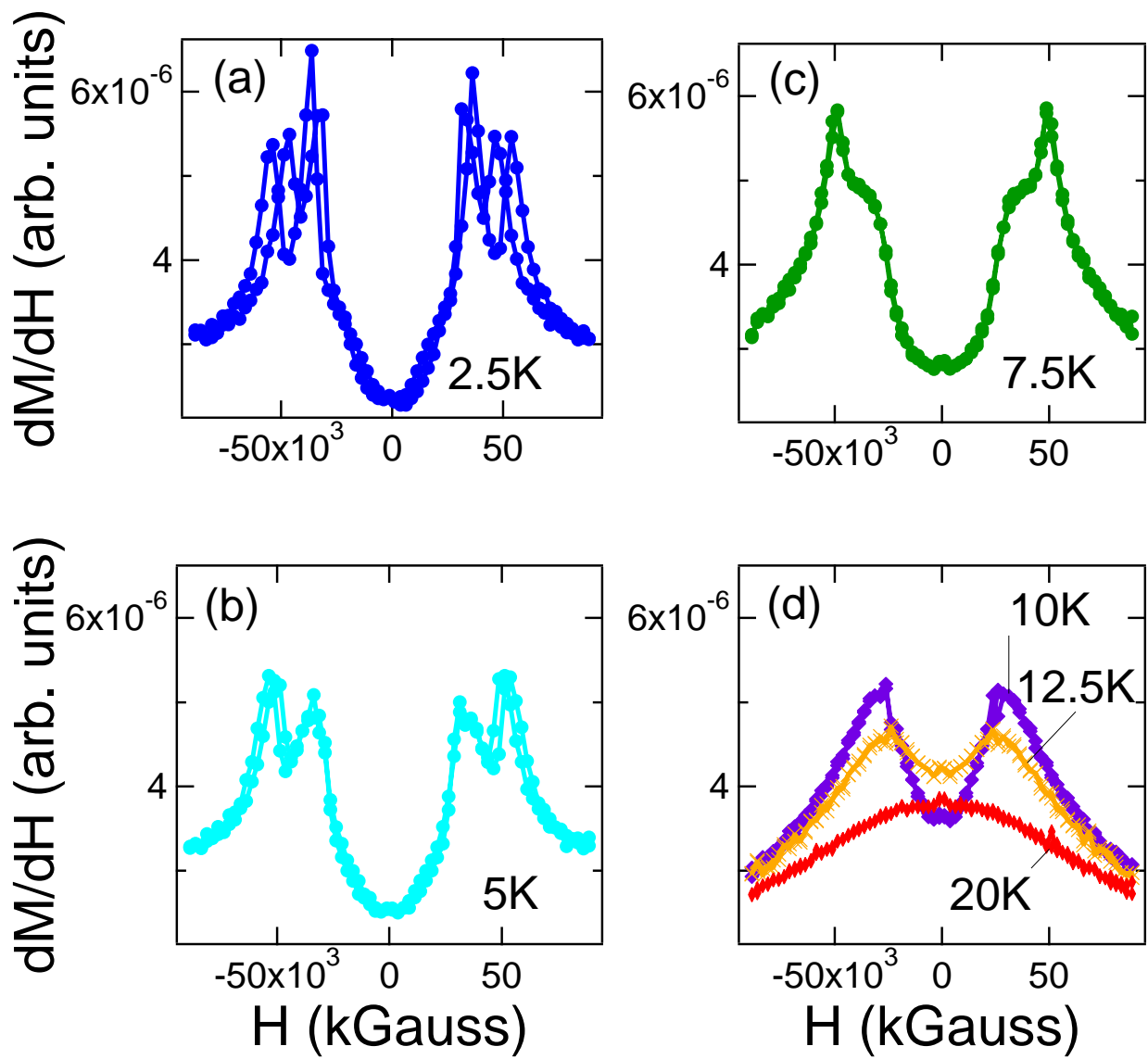}
\caption{Field dependence of dM/dH at (a) 2.5K, (b) 5K, (c) 7.5K,
(d) 10, 12.5, and 20 K. }
 \label{DMDH}
\end{center} \end{figure}

The two features in the M vs. H data can be more clearly resolved
by plotting the differential susceptibility, dM/dH, shown in fig.
\ref{DMDH}. The two maxima in the derivative plots mark the
location of the low field and high field magnetic phase
transitions.  The positive and negative field portions of the
plots very accurately mirror each other, confirming the quality of
the data. The close proximity of the low field and high field
transitions at 7.5K make it impossible to precisely locate the
center of the low field transition. In the lowest temperatures
(2.5 and 5K), the location of both transitions depends on the
measurement history. At these temperatures, there is an upper and
a lower phase boundary for both the high field and the low field
transition due to the observed hysteresis.

\begin{figure} \begin{center}
\includegraphics[width=7in]{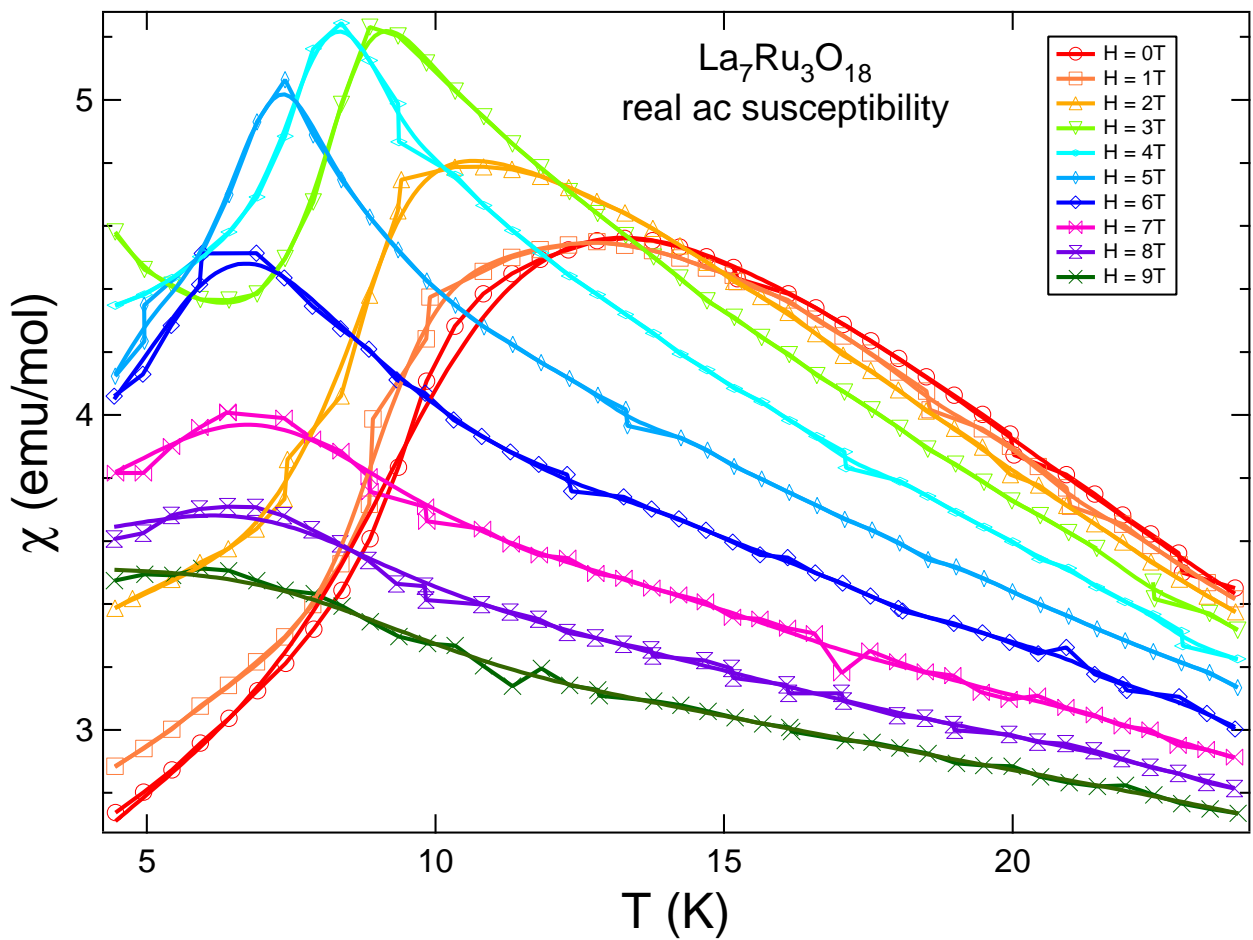}
\caption{The real portion of the ac susceptibility
($\chi\prime_{ac}$) of La$_7$Ru$_3$O$_{18}$.  Solid lines are a
guide to the eye.}
 \label{HIGHAC}
\end{center} \end{figure}

A second, complimentary means of observing the magnetic phase
boundaries is measurement of the ac susceptibility, $\chi_{ac}$.
Measurements of $\chi_{ac}$ can scan temperature while keeping the
applied field fixed, while M vs H scans fix the temperature and
vary the field.  Unlike the dM/dH plots, the $\chi_{ac}$ data can
accurately map the phase boundaries near the zero-field transition
temperature. The results of the ac susceptibility measurements on
La$_7$Ru$_3$O$_{18}$ are shown in fig. \ref{HIGHAC}.

Peaks are observed in two temperature ranges in the $\chi_{ac}$
scans.  Figure \ref{HIGHAC} shows the higher temperature peak.  It
can be seen that the shape of this peak changes from broad and
asymmetric to sharp and more symmetric as the field is increased
from 0T to 5T.  Above 5T the intensity of the peak rapidly
decreases.  It can be seen that a low-temperature peak is present
for the 3T data.  A low temperature peak is found only within a
narrow window ($\sim$0.25T) of applied fields near 3T (data not
shown). The location of the peak shifts to lower temperatures with
increasing field.  The very low temperature of this second ac
transition ($<$4K) prevented a detailed determination of the low
temperature phase boundary, although one data point could be
plotted in the overall magnetic phase diagram.

\begin{figure} \begin{center}
\includegraphics[width=7in]{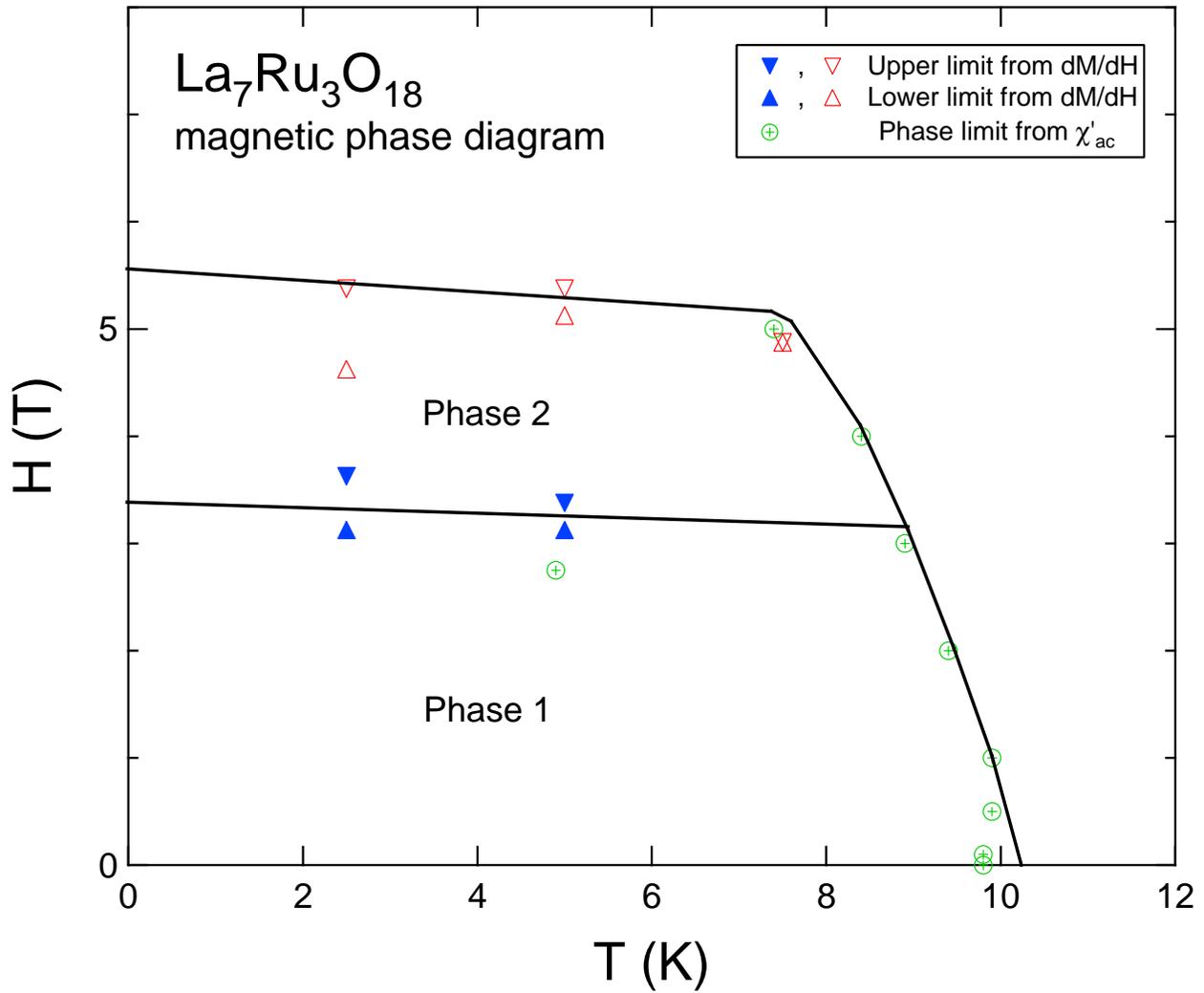}
\caption{Magnetic phase diagram of La$_7$Ru$_3$O$_{18}$.  Solid
lines are a guide to the eye.}
 \label{PD}
\end{center} \end{figure}

Combining the dM/dH and the $\chi_{ac}$ data allows the magnetic
phase diagram of La$_7$Ru$_3$O$_{18}$ to be drawn (fig. \ref{PD}).
Two definite magnetic phases are labelled in the diagram. The
dM/dH data clearly define the horizontal field boundaries between
the different phases, while the ac susceptibility data provide the
best measure of the transition temperatures of the phases.  We
note that for ac susceptibility measurements with applied fields
of $\leq$ 2T, the transition temperature was taken to be the point
of maximum slope in $\chi_{ac}$ vs. T rather than the peak maximum
due to the broad shape of the peak at low fields. Even though the
highest field (H $\geq$ 6T) measurements of $\chi_{ac}$ still show
a peak maximum, these points were not plotted on the phase diagram
due to the broader and weaker character of the peaks.  Similarly,
the peaks observed in dM/dH plots for T $\geq$ 10K were no longer
sharp and were not plotted in the phase diagram.  The observation
of weak features outside the plotted limits of the magnetic phase
diagram reflects the difficulty of defining of sharp boundaries
for magnetic phase transition, which often involve continuous and
gradual spin rearrangements in addition to exhibiting short range
correlations at temperatures just above the onset of long-range
order.

\section{Conclusions}

Since the both the low field (3T) and high field transition (5T)
involve a jump in the magnetization due to the increasing magnetic
field, these transitions should involve spin reorientation, with
spins moving from an axis defined by the crystal fields or the
antiferromagnetic interactions to an axis defined by the magnetic
field. From previously published structural data \cite{mypaper3},
it is known that there are two different types of layers in
La$_7$Ru$_3$O$_{18}$.  The thicker layer has a 2.67:1 La:Ru ratio
while the thinner layer has a 2:1 La:Ru ratio and has two
inequivalent Ru sites. The different crystal environments may give
rise to crystals fields of different strengths on the different Ru
spins, with the effect that different magnetic field strengths are
necessary to dislodge the spins from the crystal field in each
type of Ru site.  It should be noted that since $\mu g H/k$ is 18
K for an applied field of 9T, the energy scale of the applied
field is exceeds that of the antiferromagnetic ordering (T$_N
\sim$ 10 K), where $\mu$ is the moment of $S$ = 3/2 Ru, The
$g$-factor $\cong$ 2, H is the applied field, and $k$ is the
Boltzmann constant. Even though our data is from a powder sample,
it is unlikely that the two transitions represent two different
orientations of the grains in our powder since no evidence of
preferred orientation was observed in previous structural studies
\cite{mypaper3}. Although it is interesting to speculate on the
nature of the different magnetic phases, it is difficult to make
definite conclusions about the precise nature of the magnetic
ordering in the absence of neutron diffraction data or single
crystal magnetization measurements.

%Mention the change in chi_ac peak shape above and below 3T.

%There is not really paramagnetism above 5T -- still have a peak in chi_dc!!

\section{Acknowledgement}

This research was supported by the NSF Solid State Chemistry and
Polymers program, grant No. DMR-9725979.  Thanks go to B. Sales
for insightful comments on the manuscript.

%%%%%%%%%%%%%%%%%%%%%%%%%%%%%%%  bibliography

\bibliographystyle{prsty}

%%%%%%%%%%%%%%%%%%%%%%%%%%%%%%%  captions

\newpage

%\section*{Figure Captions}

%\newpage
%
%\begin{figure} \begin{center}
%\includegraphics{fig3-clean.eps}
%\caption{Top: DOS plot for La$_3$Ru$_3$O$_{11}$. Bottom: DOS plot
%for La$_4$Ru$_6$O$_{19}$.  The Fermi energy is at 0 eV in both
%plots.} \label{DOS}
%\end{center} \end{figure}

%---------Figure dimensions-----------
%RUTETRA - 5.862 x 4.525(PHOTOSHOP)
%CHIDC - 9.083 x 6.583 in (IGOR)
%MVSH - 8.097 x 6.583 (IGOR)
%DMDH - 7.514 x 6.472 (IGOR)
%HIGHAC - 9.083 x 6.583 (IGOR)
%PD - 8.097 x 6.583 (PHOTOSHOP)
%-------------------------------------

\end{document}